\documentclass[10pt, twocolumn]{IEEEtran}
 \makeatletter

\newcommand{\Rmnum}[1]{\expandafter\@slowromancap\romannumeral #1@}
\makeatother
\usepackage{cleveref}
\usepackage{graphicx}
\usepackage{subfigmat}
\usepackage{subfigure}
\usepackage{etoolbox}
\usepackage{graphicx,cite,epsfig,amssymb,amsmath,multirow}
\usepackage{graphicx,amsmath,amssymb,amsfonts,cite}
\usepackage{subfigure}
\usepackage{subfigmat}
\usepackage{etoolbox}
\usepackage{mathrsfs}
\usepackage{algorithm}
\usepackage{algorithmic}
\usepackage{float}
\usepackage{bm}
\usepackage{ulem}
\usepackage{color}
\usepackage{algorithm}
\usepackage{algorithmic}
\renewcommand{\algorithmicrequire}{\textbf{Input:}}
\renewcommand{\algorithmicensure}{\textbf{Output:}}
\DeclareMathOperator*{\argmax}{argmax}
\DeclareMathOperator*{\argmin}{argmin}

\makeatother
\begin{document}
\bibliographystyle{ieeetr}


\title{\textcolor{black}{Beam Selection for MmWave Massive MIMO Systems Under Hybrid Transceiver Architecture}}
\author{Shuang Qiu, Kai Luo, and Tao Jiang\\
\thanks{
S. Qiu, K. Luo and T. Jiang are with the School
of Electronic Information and Communications, Huazhong University of
Science and Technology, Wuhan 430074, P. R. China. 
}}

\markboth{ } {{et al.}:  }
\maketitle

\begin{abstract}
Hybrid analog/digital precoding is a promising technology to cut down hardware cost in massive multi-input multi-output (MIMO) systems at millimeter wave frequencies.
In this letter, we propose a joint beam selection scheme for analog domain precoding under discrete lens array.
By considering channel correlation among users, the proposed scheme is able to avoid inter-user interference and maximize system sum rate.
Moreover, a bio-inspired ant colony optimization-based algorithm is proposed to obtain a near-optimal solution with dramatically reduced computational complexity.
Finally, simulations show the advantages of the proposed scheme in improving system sum~rate.
\end{abstract}


%
\IEEEpeerreviewmaketitle

\section{Introduction}
Recently, the combination of massive multi-input multi-output (MIMO) and millimeter wave (mmWave) has drawn lots of attentions due to the capacity of increasing system sum~rate~\cite{SwindSep2014}.
However, due to the large scale number of base station~(BS) antennas,
mmWave massive MIMO systems suffer prohibitively high energy consumption and hardware cost, especially the cost by radio frequency (RF) chains~\cite{liangwcl2014}.

To reduce the required number of RF chains, hybrid analog/digital transceiver architectures are proposed.
One of them is mmWave beamspace massive MIMO architecture which is introduced by the concept of beamspace MIMO and a discrete lens array~(DLA) in analog domain~\cite{SayeedBrady9-13Dec.2013}.
Under this architecture, conventional antenna space is transformed into beamspace and \textcolor{black}{BSs} produce \textcolor{black}{$N$ pre-defined beam directions~($N$ is the number of BS antennas)}.
Due to channel sparsity~\cite{SwindSep2014}, only a few beams contribute to system performance.
Therefore, efficient beam selection is essential to reduce the number of RF chains and enhance system~performance.

Conventional schemes select beams by \textcolor{black}{maximal magnitude (MM)} or a threshold-based magnitude~\cite{SayeedBrady9-13Dec.2013}.
These schemes have low computational complexity, while ignore the inter-user interference which is the main factor limiting the performance of multi-user systems~\cite{QiuTVT2017}.
Due to the limited number of pre-defined beams, the users who have similar channels might be assigned identical beams and suffer severe inter-user interference.
To overcome this problem, an interference-aware (IA) scheme was proposed to reselect beams for the interference users~\cite{GaoDaiChenEtAlMay2016}.
Moreover, beam selection based on the criteria of signal-to-interference-ratio and capacity maximization were also proposed~\cite{AmadoriJune2015}.
\textcolor{black}{However, the existing schemes are all in a greedy way and inter-user interference is not fully considered.}
Furthermore, since the users tend to be closely located with the number of smart devices increasing, the channel correlation between users increases and deteriorates inter-user interference.
Therefore, jointly selecting beams for all users is vital to enhance the overall system performance.

\begin{figure*}
\centering
\includegraphics[width=5.0in]{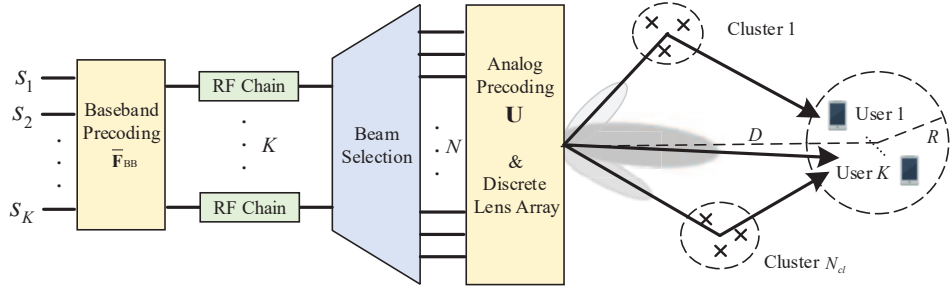}
\caption{The transceiver of mmWave beamspace massive MIMO with $K$ RF chains and $K$ closely-located users.}
\label{system_model_new}
\end{figure*}


In this letter, a joint beam selection scheme is proposed by fully considering channel correlation among users.
We aim to maximize system sum rate to approach the performance of fully digital precoding.
However, finding the optimal beams requires an exhaustive search which is computationally prohibitive.
Therefore, we innovatively take the beam selection problem as a traveling problem and the idea of ant colony optimization (ACO) is utilized to obtain a near-optimal solution.
As a typical bio-inspired optimization algorithm in solving discrete optimization problems, ACO has been widely used in various domains~\cite{DorigoNov.2006}, such as coverage problem in wireless sensor networks~\cite{LeeTII2011}, detection in MIMO systems~\cite{LainChenAugust2010} etc.
By proposing the ACO-based beam selection algorithm, a near-optimal beam set is obtained with dramatically reduced computational complexity.
Finally, simulation and theoretical analysis results show that the proposed beam selection scheme outperforms the existing ones in improving system sum rate, especially when the users have high channel~correlation.
Moreover, near-identical system sum rate to the exhaustive search could be achieved.


\section{System and Channel Models}\label{section:model}

A downlink mmWave beamspace massive MIMO system is considered where the BS is equipped with~$N$ antennas serving~$K$ single-antenna users (see Fig.~\ref{system_model_new}).
\textcolor{black}{To model the scenario where users are densely located wtih high channel correlation, the users are assumed to be randomly located inside a ring of radius~$R$ at a distance~$D$ away from the BS.}
\textcolor{black}{Moreover, $K$ RF chains are considered in analog domain}.

When the DLA is employed, conventional antenna space can be expressed as beamspace through a spatial Fourier transform matrix $\mathbf{U}$ in analog domain.
This transform matrix includes a set of orthogonal basis and is given as
\vspace{-0.3em}
\begin{equation}\label{20170318 eq1}
\mathbf{U} =\left[ {\begin{array}{*{5}{c}}
{\begin{array}{*{5}{c}}
{\hspace{-0.75em}\mathbf{a}\left( {{\theta _1}} \right),}&{ \cdots ,}
\end{array}}&{\hspace{-0.75em}\mathbf{a}\left( {{\theta _{N }}} \right)}
\end{array}} \right],
\end{equation}
where $\mathbf{a}\left( \theta_n \right), n=1,\dots,N,$ denotes a steering vector given~as
\vspace{-0.7em}
\begin{equation}\label{20170317 eq3}
\mathbf{a}\left( \theta_n \right)= \frac{1}{\sqrt{N}} \left[ e^{-j2 \pi \theta_n m }  \right]_{m \in  \mathcal{J }\left( N \right)},
\end{equation}
where $\theta_n = \frac{1}{N}\left( n-\frac{N+1}{2} \right), n=1,\dots,N,$ are pre-defined spatial directions and $\mathcal{J }\left( N \right)= \left\{ q-\frac{N+1}{2}, q=1, \dots, N \right\}$.

Thus, the downlink received signal of the $K$ users is
\vspace{-0.3em}
\begin{equation}\label{20170318 eq2}
{ \mathbf{y }^{\mathrm{DL}}} = \textcolor{black}{ {\mathbf{H}^H}{\mathbf{U}}}{\mathbf{F}_{\mathrm{BB}}}\mathbf{x} + \mathbf{n}= \textcolor{black}{{\overline {\mathbf{H} }^H}}{\mathbf{F}_{\mathrm{BB}} }\mathbf{x} + \mathbf{n},
\end{equation}
where $\mathbf{x} \in \mathbb{C}^{K \times 1}$ is transmitted complex signal vector to the~$K$ {users} satisfying $\mathrm{E}\left\{ \mathbf{x} \mathbf{x}^H  \right\}=\mathbf{I}_K$, $\mathbf{H} \in \mathbb{C}^{N \times K}$ is channel matrix, the vector $\mathbf{n} \in \mathbb{C}^{K \times 1}$ represents the additive white Gaussian noise whose entries are zero-mean and variance $\sigma ^2$, $\mathbf{F} _{\mathrm{BB}}\in \mathbb{C}^{N \times K}$ denotes digital precoding matrix and~${\overline {\mathbf{H} }}=\mathbf{U}^H\mathbf{H}$ is beamspace representation of~$\mathbf{H}$.

Let $N_{cl}$ be the number of multipath clusters and each cluster is composed of $N_{ray}$ subpaths.
Then, the channel vector for the $k$-th user could be modeled by light of sight (LoS) and non-LoS (NLoS) components as~\cite{liangwcl2014}
\vspace{-0.4em}
\begin{align}\label{20170319 eq2}
 \mathbf{h}_k
 =\alpha_{k,0}  \mathbf{a} \left(\phi_{k,0} \right)  + \sum_{i=1}^{N_{cl}}  \sum_{l=1}^{N_{ray}}  \alpha_{k,i,l} \mathbf{a} \left(\phi_{k,i,l}\right),
\end{align}
where $\alpha_{k,0} $ and $\alpha_{k,i,l}$ present the complex gains of LoS and NLoS subpaths, respectively, the parameters~$\phi_{k,0}$ and~$\phi_{k,i,l}$ denote the corresponding spatial directions defined as~$\phi =\frac{d}{\lambda} \mathrm{sin} \omega$ with $\lambda$ denoting wavelength, $d$ antenna spacing given as $\frac{\lambda}{2}$ and $\omega$ the spatial angle of arrival (AoA).

Moreover, the beamspace channel matrix is
\begin{equation}\label{20170319 eq1}
{\overline {\mathbf{H} }}=\left[ \mathbf{U}^H \mathbf{h}_1, \dots, \mathbf{U}^H \mathbf{h}_K  \right]=\left[  \overline{ \mathbf{h}}_1, \dots,  \overline{\mathbf{h}}_K  \right],
\end{equation}
where $\overline{\mathbf{h}}_k, k=1, \dots, K,$ is the beamspace channel vector of the~$k$-th user.
The elements of~$\overline{\mathbf{h}}_k$ present the channel gains provided by the $N$ pre-defined beams.
Due to channel sparsity in mmWave systems, only several dominant elements exist in~$\overline{\mathbf{h}}_k$ counting for the transmission rate of the $k$-th user.
When the BS is equipped with $K$ RF chains, a reduced-dimension matrix~$\overline {\mathbf{H} } _s\in \mathbb{C}^{K \times K} = \overline {\mathbf{H} } \left( m,:   \right)_{m \in \mathcal{S} }$ should be reconstructed, where~$\mathcal{S}= \left\{ I_1,\dots,I_K   \right\} $ is the index set of the selected beams.
Thus, the downlink received signal of the $K$ users is rewritten~as
\begin{equation}\label{20170319 eq4}
{ \mathbf{y}_s ^{\mathrm{DL}}}
= \textcolor{black}{{\overline {\mathbf{H} }_s^H}}{\overline{\mathbf{F}}_\mathrm{BB}}\mathbf{x} + \mathbf{n},
\end{equation}
where $\overline{\mathbf{F}}_\mathrm{BB}\in \mathbb{C}^{K \times K} $ is the reduced-dimension digital precoding matrix based on~$\overline {\mathbf{H} }_s $.

\textcolor{black}{In the mmWave beamspace massive MIMO system, the ideal performance can only be obtained when the users located in orthogonal subspace.
However, user channels are usually correlated in real propagation environment.
Therefore, advanced beam selection scheme should take channel correlation into consideration to maximize system performance.}
\section{The Proposed Joint Beam Selection Scheme}\label{section:algorithm}


\subsection{Problem Formulation}\label{subsection:formulation}
Since we mainly focus on the beam selection in analog domain, a widely used zero-forcing~(ZF) precoder is adopted for baseband and given as
$\overline{\mathbf{F}}_\mathrm{BB} = \alpha \overline {\mathbf{H} }_s \left(\overline {\mathbf{H} }_s ^H \overline {\mathbf{H} }_s \right)^{-1}$,
where~$\alpha$ is a scaling factor to ensure $\mathrm{E}\left\{  \left\| \overline{\mathbf{F}}_\mathrm{BB} \mathbf{x}   \right\|^2_2     \right\}=\rho$ and~$\rho$ is transmit power of BS.
Thus, the factor~$\alpha$ is given as
\begin{equation}\label{20170319 eq6}
\alpha = \sqrt{\frac{\rho}{ \mathrm{tr}\left( \left(\overline {\mathbf{H} }_s ^H \overline {\mathbf{H} }_s \right)^{-1}   \right)    }    }.
\end{equation}
When equal power allocation scheme is employed at BS, the average rate of the $k$-th user is obtained as
\vspace{-0.3em}
\begin{align}\label{20170319 eq7}
R_k  &= \mathrm{log}_2\left( 1+ \frac{ \left| \alpha \right|^2}{\sigma^2 K}   \right).
\end{align}
Then, the system sum rate is
$R^{\mathrm{sum}}=\sum_{k=1}^K R_k$.
By exploiting the channel sparsity, the optimal beams to maximize the system sum rate should be selected by
\begin{equation}\label{20170319 eq8.5}
\overline {\mathbf{H} }_s^{\mathrm{opt}} =\mathop {\argmax }\limits_\mathcal{S}  \; R^{\mathrm{sum}}.
\end{equation}
Based on (\ref{20170319 eq6})--(\ref{20170319 eq8.5}), the matrix $\overline {\mathbf{H} }_s^{\mathrm{opt}}$ is further formulated as
\vspace{-0.3em}
\begin{equation}\label{20170319 eq9}
\overline {\mathbf{H} }_s^{\mathrm{opt}} = \mathop {\argmin }\limits_\mathcal{S} \; \mathrm{tr}\left(\hspace{-0.2em} \left(\overline {\mathbf{H} } \left( m,:   \right)_{m \in \mathcal{S} }^H \overline {\mathbf{H} } \left( m,:   \right)_{m \in \mathcal{S} } \right)^{-1}  \hspace{-0.2em} \right) .
\end{equation}
The optimal solution to obtain $\overline {\mathbf{H} }_s^{\mathrm{opt}}$ is exhaustive search, while exhaustive search suffers prohibitively high computational complexity with~$\left( \hspace{-0.3em} {\begin{array}{*{20}{c}}
N\\
K
\end{array}} \hspace{-0.3em}\right)$ possible combinations.
Since low-complexity precoding is essential in massive MIMO systems~\cite{XiongWCM2749}, a low-complexity algorithm for beam selection is required to obtain a near-optimal solution.

\subsection{\textcolor{black}{ACO-Based Beam Selection Algorithm}}\label{subsection:algorithm}

We creatively take the beam selection as a traveling problem, where the users are regarded as cities and the pre-defined beams are taken as optional paths for each user.
Our goal is to travel all the cities~($K$ users) once with~$K$ paths~($K$ beams) under lowest cost~(criterion given in (\ref{20170319 eq9})).
Inspired by ACO, a low-complexity iterative algorithm is proposed using a positive feedback mechanism
to achieve a near-optimal beam selection.

Assume that there are $T_{\mathrm{max}}$ iterations in the algorithm.
During the $t$-th iteration, there are $N$ possible beam selections for the $k$-th user.
However, since the number of dominant beams is much smaller than~$N$ due to channel sparsity, we could only use~$B_k$ dominant beams\footnote{The value of $B_k, \forall k,$ could be pre-defined as a fixed constant or be obtained by setting thresholds on the magnitudes of beams.} for the beam selection of the $k$-th user. \textcolor{black}{The index set} is denoted as $\mathcal{B}_k=\{I_{k1},\dots,I_{kB_k}  \}$.
Then, a utility function~$d_{kb}^t$ is defined based on~(\ref{20170319 eq9}) to evaluate the suitability of the~$b$-th beam in~$\mathcal{B}_k$ and given as
\vspace{-0.6em}
\begin{equation}\label{20170322 eq1}
\textcolor{black}{d_{kb}^t= \mathrm{tr}\left( \left( ( \overline{\underline{{\mathbf{H} }}}_{s,k}^{t-1} )^H \overline{\underline{{\mathbf{H} }}}_{s,k}^{t-1}    +\widehat{\mathbf{H}} \right)^{-1}   \right),}
\end{equation}
\textcolor{black}{where $\overline{\underline{{\mathbf{H} }}}_{s,k}^{t-1} \in \mathbb{C} ^{(K-1) \times K}$ is obtained by removing the~$k$-th row of $\overline{{{\mathbf{H} }}}_s^{t-1}$ with $\overline{{{\mathbf{H} }}}_s^{t-1}$ denoting the channel matrix obtained at the $t-1$ iteration, and $ \widehat{\mathbf{H}} = \overline {\mathbf{h}} {(I_{kb},:)^H} \overline {\mathbf{h}} (I_{kb},:)+\varsigma \mathbf{I}$ where~$\overline {\mathbf{h}} (I_{kb},:)$ is the $I_{kb}$-th row of $\overline {\mathbf{H}}$ and $\varsigma$ is a small positive number to guarantee that the matrix inversion~exists.}

Inspired by the ACO~\cite{DorigoNov.2006}, the beam for the~$k$-th user is selected according to a probability~$p_{kb}^t$, \textcolor{black}{which is not only decided by utility function~$d_{kb}^t$, but also by a positive feedback factor~$\tau^{t-1}_{kb}$ of the previous selection result.
The positive feedback factor is used to adjust the impact of~$d_{kb}^t$ on beam selection from the perspective of overall system~performance.}

Firstly, a parameter~$\eta_{kb}^t$ is defined to denote the impact of~$d_{kb}^t$ on beam selection probability.
Since the BS prefers to select the beam with smaller $d_{kb}^t$,
the parameter~$\eta_{kb}^t$ should be inversely proportional to~$d_{kb}^t$ and given as
\vspace{-0.7em}
\begin{equation}\label{20170322 eq2}
\eta_{kb}^t \propto \frac{1}{d_{kb}^t}.
\end{equation}

Then, the BS calculates the selection probability of each beam based on both $\eta_{kb}^t$ and $\tau^{t-1}_{kb}$, which is given as
\vspace{-0.5em}
\begin{equation}\label{20170322 eq3}
p_{kb}^t=\frac{[\tau_{kb}^{t-1}]^{a}[\eta_{kb}^t]^q}{\sum_{b=1}^{B_k} [\tau_{kb}^{t-1}]^{a}[\eta_{kb}^t]^q},
\end{equation}
where
$a$ and $q$ are positive constants denoting the weights of~$\tau_{kb}^{t-1}$ and $ \eta _{kb}^t$ on beam selection, respectively.
The beam with the maximal probability ${p_{kb}^t}, b=1,\dots, B_k,$ is temporarily selected for the~$k$-th user.
\textcolor{black}{In (\ref{20170322 eq3}), $ \eta _{kb}^t$ can be taken as a priori information for beam selection.
When $\tau_{kb}^{t-1}$ is ignored or~$a=0$, the proposed algorithm reduces to a classic greedy algorithm and only a locally optimal solution can be obtained.
The positive feedback $\tau_{kb}^{t-1}$ adjusts the impact of $ \eta _{kb}^t$ and makes it possible for any beam to be selected.
Therefore, an approximate globally optimal solution can be found.}

Finally, the positive feedback factor will be updated after the beam selection.
It is influenced by the selection probability ${p_{kb}^t}$ at the $t$-th iteration and parameter $\eta_{kb}^{t}$. Thus, $\tau_{kb}^{t}$ is given as
\vspace{-0.5em}
\begin{equation}\label{20170322 eq4}
\tau_{kb}^{t}=\left(1-\gamma \right)\tau_{kb}^{t-1}+ \omega \eta_{kb}^{t} p_{kb}^{t},
\end{equation}
\textcolor{black}{where~$\omega $ controls the increasement of positive feedback factor during the $t$-th iteration and $\gamma \in (0,1)$ is a decay parameter to control
the convergence rate of the algorithm.}
\textcolor{black}{
Increasing~$\gamma$ weakens the influence of the previous positive feedback.
The beam with higher probability $ p_{kb}^{t}$ and~$\eta_{kb}^{t}$ gains more increasement of positive feedback, which can accelerate the update of beam selection.
Therefore, near optimal~beam indices can be found faster with $T_{\mathrm{max}}$ increasing.}
In summary, the ACO-based beam selection algorithm is given in~Algorithm~\ref{algaco-based}.

\begin{algorithm}[!t]
    \renewcommand{\algorithmicrequire}{\textbf{Input:}}
	\renewcommand{\algorithmicensure}{\textbf{Output:}}
	\caption{\textcolor{black}{ACO-based beam selection algorithm}}
    \label{algaco-based}
	\begin{algorithmic}[1]
        \REQUIRE 
        $\overline{\mathbf{H}}$, \textcolor{black}{$B_k, \forall k$,} $T_{\mathrm{max}}$, $a$, $b$, $\gamma$, $\omega$
		\ENSURE $\overline{\mathbf{H}}_s$
      			\STATE \textbf{Initialize} $\mathcal{S}^0=\left\{I_1^0, \dots,  I_K^0 \right\}={\mathcal{S}}_{\mathrm{MM}}$ \;(The set of beam indices with the largest magnitude for each user~\cite{SayeedBrady9-13Dec.2013});
      $\overline{\mathbf{H}}_s^{0} = \overline{\mathbf{H}}(m,:)_{m \in \mathcal{S}^{0}}$;
      $\mathrm{Trace}=\infty$; $\eta_{kb}^t =1, \forall t,k,b$.\\
\FOR {$t = 1: T_{\mathrm{max}}$}
               \STATE $\mathcal{S}^t=\mathcal{S}^{t-1};$
              \FOR {$k = 1: K$}
                \FOR {$b = 1: B_k$}
      			\STATE Compute $d_{kb}^t$, $\eta_{kb}^t$, $p_{kb}^t$ according to (\ref{20170322 eq1}), (\ref{20170322 eq2}) and~(\ref{20170322 eq3}), respectively;
      			\STATE Update $\tau_{kb}^t$ according to (\ref{20170322 eq4});
                \ENDFOR
      			\STATE Find the maximal $p_{kb}^t, \forall b$ and record its index $b_{\mathrm{max}}$;
      			\STATE $I_k^{t-1}=$\textcolor{black}{$I_{kb_{\mathrm{max}}}$};		\IF {$d_{k b_{\mathrm{max}}}^t \leq \mathrm{Trace}$}
      \STATE $\mathrm{Trace}=d_{kb_{\mathrm{max}}}^t$, $\overline{\mathbf{H}}^t_s = \overline{\mathbf{H}}(m,:)_{m \in \mathcal{S}^{t-1}}$;
      \ELSE
      \STATE $\overline{\mathbf{H}}^t_s = \overline{\mathbf{H}}(m,:)_{m \in \mathcal{S}^{t}}$;
      \ENDIF
      \ENDFOR
       		\ENDFOR	
       \STATE Return $\overline{\mathbf{H}}_s$
    \end{algorithmic}
\end{algorithm}

\subsection{Computational Complexity Analysis}\label{compl}
The computational complexity of the proposed beam selection scheme is compared to the existing ones, i.e., \textcolor{black}{MM-1~\cite{SayeedBrady9-13Dec.2013}} and IA algorithm \cite{GaoDaiChenEtAlMay2016}.
\textcolor{black}{The main complexity of IA algorithm comes from reselecting beams for interference users~(IUs) who have identical beam indices with largest magnitude.
IA algorithm selects different beams for each user one by one.
When~$\overline{K}$ IUs exist, there are~$(N-K)+(\overline{K}+1-\overline{k}), \overline{k}=1,\dots,\overline{K},$ possible beam selections for the~$\overline{k}$-th IU.}
Since~$K \times K$--dimension matrix inversion has highest complexity~($\mathcal{O}\left( K^3 \right)$) in the IA and the proposed algorithms, their complexities are compared by counting the times of performing matrix inversion.
Set~$B^{\mathrm{tol}} = \sum_{k=1}^K B_k$
Then, the computational complexity is compared in Table \ref{comp_com}.
\begin{table}[h]
\caption{\textcolor{black}{Computational complexity comparison}}\label{comp_com}
\centering
\begin{tabular}{|c|c|c|c|}
\hline
 Scheme & MM-1 &  IA &   ACO-based     \\ \hline \hline
Times & 0 & $\left( N-K \right)\overline{K}+{\left( \overline{K}^2+\overline{K} \right)}/{2}$ & $T_{\mathrm{max}} B^{\mathrm{tol}}$ \\ \hline
\end{tabular}
\end{table}


When users have higher channel correlations, $\overline{K}$ becomes larger and the IA algorithm suffers more computational cost.
Fortunately, the computational complexity of the proposed algorithm is unrelated to $\overline{K}$, which makes it more applicable for the scenario with high channel correlation.

%
\begin{figure*}
\begin{subfigmatrix}{3}
\subfigure[ Sum rate comparison versus different transmit powers with $N=32$, $K=5$, $B_k=10, \forall k,$ and~$T_{\mathrm{max}}=10$.]{
\includegraphics[width=0.32\textwidth,keepaspectratio]{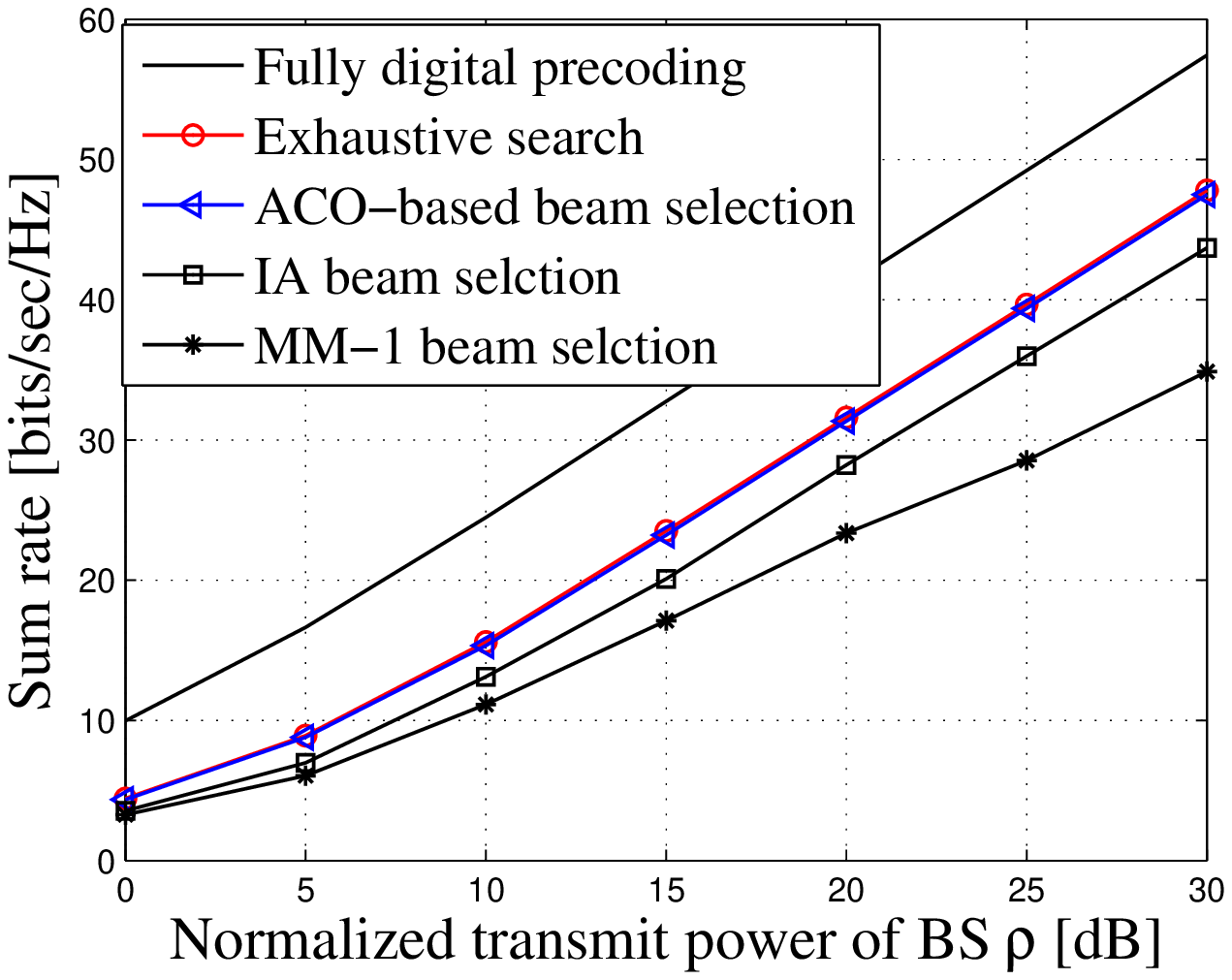}}
\subfigure[ Sum rate comparison versus different numbers of users with~$\rho=20$ dB, $N=100$, $B_k=10, \forall k$ and $T_{\mathrm{max}}=10$.]{
\includegraphics[width=0.32\textwidth,keepaspectratio]{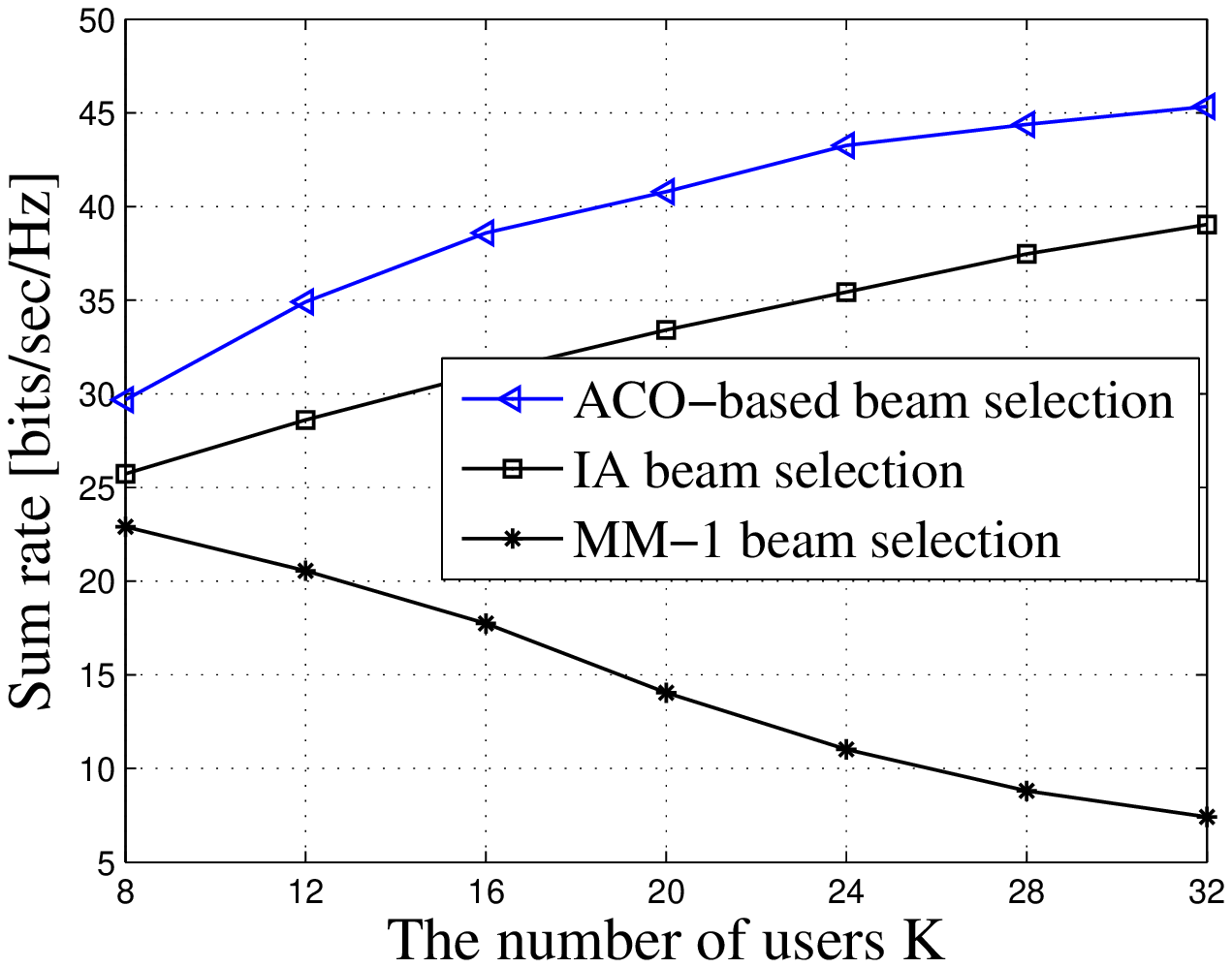}}
\subfigure[ Sum rate comparison versus different numbers of~$B_k, \forall k,$ ($T_{\mathrm{max}}=10$) and~$T_{\mathrm{max}}$ ($B_k=10, \forall k,$) with~$\rho=20$ dB, $N=100$, $K=16$.]{
\includegraphics[width=0.32\textwidth,keepaspectratio]{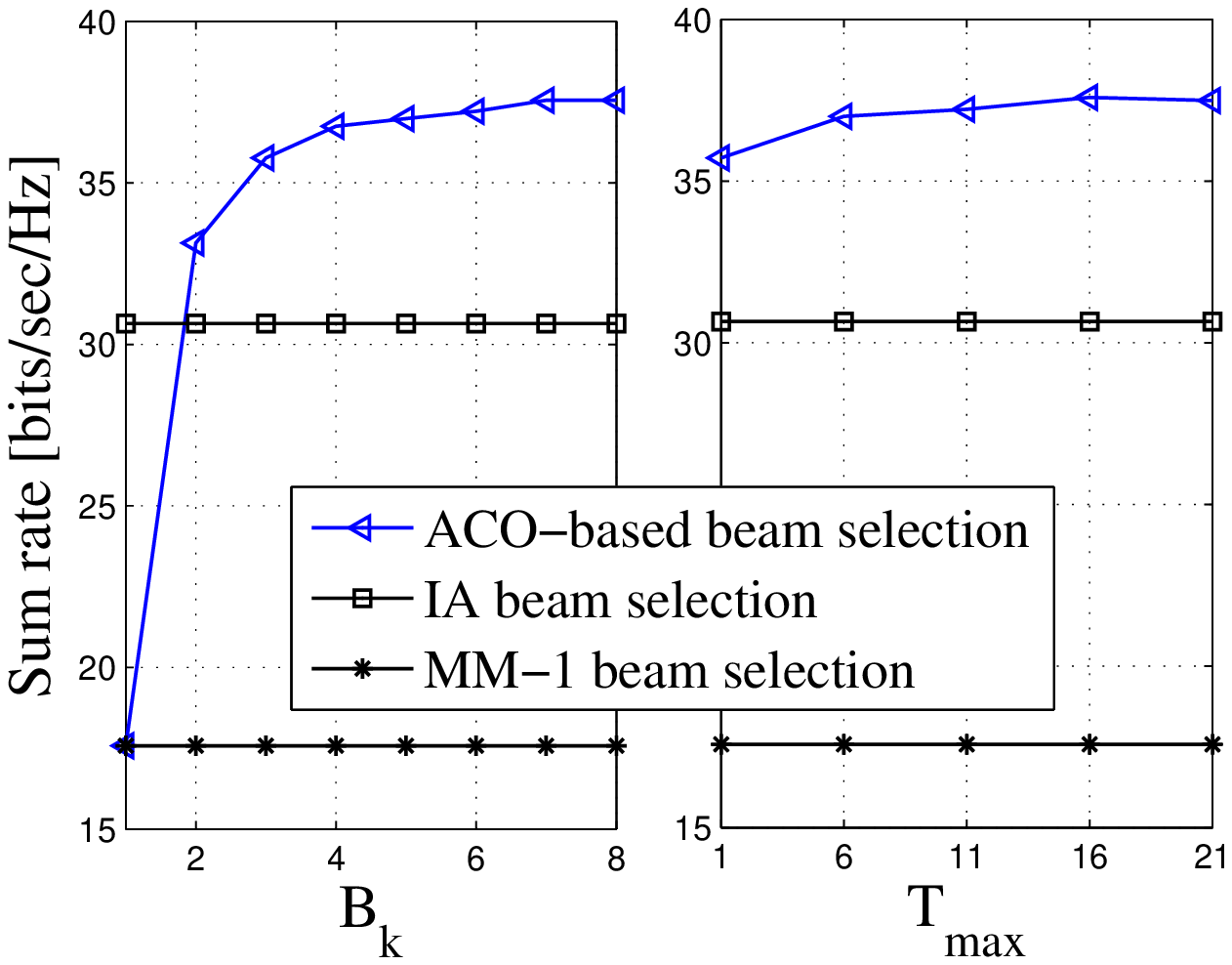}}
\end{subfigmatrix}
\caption{System sum rate comparisons of different beam selection schemes.}
\label{simulations}
\end{figure*}

\section{Simulation Results and Discussions}\label{section:simulation}
In this section, the performance of the proposed scheme is evaluated.
The AoAs~$\phi_{k,0},\forall k,$ of the LoS paths are assumed to be~uniformly distributed as $\phi_{k,0} \sim \mathcal{U}\left[ \omega_0, \omega_0+2\mathrm{arcsin}\left( \frac{R}{D}  \right) \right] $ with~$\omega_0 \in \left[ -\frac{\pi}{2},\frac{\pi}{2} \right] $.
Besides, the AoAs of the $N_{ray}$ subpaths in one cluster are distributed over $\left[ \overline{\varphi} - \varphi_{\Delta} /2, \overline{\varphi} + \varphi_{\Delta} /2 \right]$, where~$\overline{\varphi} \in \left[ -\frac{\pi}{2},\frac{\pi}{2} \right]  $ is the mean AoA of the $N_{ray}$ subpaths and $\varphi_{\Delta}$ is the angle spread (AS).
The parameters for the mmWave channel model and the ACO-based algorithm\footnote{\textcolor{black}{The parameters are decided by simulations due to the difficulty in theoretical analysis of ACO metaheuristic.} } are given in Table~\ref{simulation_parameter}.
Moreover, the function between $\eta$ and $d$ in~(\ref{20170322 eq2}) is given as $\eta =\sqrt{\frac{1}{e^{d /N^2}}}$  and the parameter $\varsigma=10^{-3}$ in the simulations.

\begin{table}[h]
\caption{\textcolor{black}{Simulation parameters}}\label{simulation_parameter}
\centering
\begin{tabular}{|c|c||c|c||c|c|}
\hline
Parameter & Value  &  Parameter &   Value   &  Parameter &   Value  \\ \hline
Distance $D$ &150 & $N_{cl}$ & 3&  $a$&0.8  \\ \hline
Radius $R$    &10& $N_{ray}$ & $\mathcal{U}[1,30]$& $q$&0.4     \\ \hline
AS $\varphi_{\Delta}$ &$5^{\circ}$ & $\omega$ &        0.5& $\alpha_{k,0}$ & -3 dB  \\ \hline
Decay $\gamma$ &0.3 & $\sigma^2$& 1 & $\alpha_{k,i,l}$ & -5 dB \\ \hline
\end{tabular}
\end{table}



\textcolor{black}{System sum rate} is compared in Fig. \ref{simulations} among the proposed ACO-based scheme, MM-1~\cite{SayeedBrady9-13Dec.2013} and IA~\cite{GaoDaiChenEtAlMay2016}.
Considering the prohibitive complexity of exhaustive search, we set~$N=32$ and~$K=5$ in Fig. 2(a).
It is indicated that the ACO-based scheme achieves near-identical performance to exhaustive search and outperforms the other two schemes. It is worthy to mention that the number of RF chains is significantly reduced by 6.4 times than that of fully digital precoding.

Fig. 2(b) depicts the system sum rate versus numbers of users.
It is observed that the ACO-based scheme obtains the best performance, even when the number of users grows large and the users have highly correlated channels.
Moreover, the MM-1 scheme obtains the worst performance due to the ignorance of channel correlation.

Regarding to computational complexity, Fig. 2(c) depicts the influence of optional beams~$B_k,\forall k,$ and iteration $T_{\mathrm{max}}$ to system sum rate.
It is shown that the system sum rate sharply increases when $B_k,\forall k,$ turns to~2 and $89 \%$ of the system performance is obtained.
Moreover, the parameter $T_{\mathrm{max}}$ slightly affects the system sum rate and $96 \%$ of the system performance could be achieved with only once iteration.

Therefore, even when $B_k=2,\forall k,$ and $T_{\mathrm{max}}=1$, the ACO-based scheme could obtain the system sum rate of~$32.0$ bits/sec/Hz, which is still larger than the other schemes.
Moreover, the ACO-based scheme computes matrix inversion of $T_{\mathrm{max}} B_k K =32$ times, \textcolor{black}{while the IA scheme computes~$\left( N-K \right)\overline{K}+\frac{1}{2}\left( \overline{K}^2+\overline{K} \right)= 84.5\overline{K}+0.5\overline{K}^2 \geq 85 $ times (see Subsection~\ref{compl}),} which is much higher than the computational complexity of the ACO-based algorithm.

%


\section{Conclusion}\label{section:conclusion}
\textcolor{black}{A joint beam selection scheme was proposed in mmWave massive MIMO systems with DLA.
The proposed scheme fully considers the channel correlation among users and maximizes system sum rate.
By taking the beam selection problem as a traveling problem, the ACO-based algorithm was proposed to obtain a near-optimal performance to exhaustive search with dramatically reduced computational complexity.
Finally, simulation results are provided to demonstrate the advantages of the proposed scheme, especially when users have highly correlated channels.
} 



\bibliographystyle{IEEEtran}

\end{document}